\documentclass[11pt]{article}
\usepackage{moriond2,epsfig}

\def\be{\begin{equation}}
\def\ee{\end{equation}}
\def\bea{\begin{eqnarray}}
\def\eea{\end{eqnarray}}

\def\lsim{\raise0.3ex\hbox{$<$\kern-0.75em\raise-1.1ex\hbox{$\sim$}}}
\def\gsim{\raise0.3ex\hbox{$>$\kern-0.75em\raise-1.1ex\hbox{$\sim$}}}

\def\beq{\begin{equation}}
\def\eeq{\end{equation}}
\def\bea{\begin{eqnarray}}
\def\eea{\end{eqnarray}}
\def\bq{\begin{quote}}
\def\eq{\end{quote}}

\begin{document}
\vspace*{4cm}
\title{LARGE TRANSVERSE MOMENTUM
SUPPRESSION OF $\pi^0$'s
IN $Au+Au$ AND $d+Au$ COLLISIONS
AT $\sqrt{s}=200$ GeV}

\author{A. CAPELLA$^{(a)}$, 
\underline{E. G. FERREIRO}$^{(b)}$, A. B. KAIDALOV$^{(c)}$,
D. SOUSA$^{(b)}$}

\address{$^{(a)}$ Laboratoire de Physique Th\'eorique, Universit\'e de Paris
XI, B\^atiment 210,
F-91405 Orsay Cedex, France\\
$^{(b)}$ Depto. de F\'{\i}sica de Part\'{\i}culas,
Universidade de Santiago de Compostela \\
E-15706 Santiago de Compostela, Spain\\
$^{(c)}$ Institute of Theoretical and Experimental
Physics, 117259 Moscow, Russia}

\maketitle\abstracts{We propose a model of suppression of $\pi^0$'s based on
two different effects: at low $p_T$ we take into account the shadowing 
corrections,
while at large $p_T$ the suppression is produced due to the interaction of the
large $p_T$ pion with the
dense medium created in the collision. The main features of the 
data on $AuAu$
and $dAu$ collisions 
are reproduced both at mid and at forward rapidities.}

\section{Introduction}
The experimental data on central$AuAu$ collisions taken at RHIC
show that the nuclear
modification factor, 
$R_{AA}(b,y,p_T) = \Big [ {dN^{AA} \over
dy d^2p_T}(b) \Big ] \Big / \Big [ n(b)
{dN^{pp} \over dyd^2p_T} \Big ]$, is smaller than 1
for all $p_T$. 
That means that the yield of
particles produced
in $AA$ collisions at mid-rapidities and large $p_T$ increases with
centrality much slower than the number of binary collisions $n(b)$.
This
phenomenon is particularly interesting since it is not observed in
$dAu$
collisions at RHIC at mid-rapidities.

In order to explain these data, one should take into account two kinds of
effects. On one side, at low $p_T$ this suppression is explained by the 
shadowing corrections \cite{1r}. At very high energies the shadowing effects
--nonlinear--
lead,
as $s \to \infty$,
to a saturation
 of the distributions of
partons in the colliding nuclei.
These corrections 
are important for the description of inclusive
spectra,
for particles with $p_T \sim <p_T>$. But with increasing
$p_T$ the shadowing corrections decrease and the scaling with
$n(b)$ is predicted in perturbative QCD ($R_{AA} \rightarrow 1$).

So in order to explain the data at large $p_T$, it is neccessary to consider
that for
particles with large momentum transfer,
there are in general
final state
interactions \cite{2r}
 --jet quenching or jet absorption--. These interactions
lead to an energy loss of the large $p_T$ parton
(particle) in the dense medium produced in the collision.
Hadrons loose a finite fraction of their longitudinal momentum due to
multiple scattering. 
From this point of view, it is natural to expect
that
a particle or a parton scattered at some non zero angle will also loose a
fraction of its transverse momentum due to final state
interactions with
a medium of partonic or
prehadronic nature. The scattered particle 
does not disappear
as a result of the interaction but its $p_T$ is shifted to smaller
values. Due to the steepness of the $p_T$ distribution, the effect of
suppression at large $p_T$ may
be quite large. Moreover, in this case there is also a gain of the
yield at small $p_T$ due to particles produced at larger
$p_T$ -- which have experienced a $p_T$ shift due to the interaction
with the medium.
This suppression vanishes at low $p_T$: when the $p_T$ of the
particle is close to $<p_T>$, its $p_T$ can either increase
or decrease as a result of the interaction, i.e. in average the $p_T$
shift tends to zero.
Our aim here is to describe the suppression of the
yield of pions
in a framework based on final state interactions and taking into account
shadowing and Cronin effect.

\section{The model}
Our approach contains dynamical, non linear shadowing. It is determined in
terms of diffractive cross sections and it applies to soft and hard processes.
The reduction of multplicity from shadowing corrections can be expressed by 
\beq
R_{AB}(b)= { \int d^2s f_A(s) f_B(b-s) \over T_{AB}(s)}, \ \
f_A(b)= {T_A(b) \over 1+A F(s) T_A(b)}
\nonumber
\eeq
where the function $F(s)$ represents the integral of the triple 
Pomeron cross section
over the single Pomeron
one: 
\beq
F(s) = 4 \pi \int_{y_{min}}^{y_{max}} dy {1 \over \sigma^P(s)}
{d^2 \sigma^{PPP} \over dy dt}|_{t=0}= C [ {\rm exp (y_{max})} - {\rm exp}
(y_{min})].
\nonumber
\eeq
From this formula one can observe that for a particle produced at $y=0$, with
$y_{max}={1 \over 2}
 {\rm ln}(s/m_T^2)$ and 
$y_{min}= {\rm ln}(R_A m_N/\sqrt{3})$,
the effect of shadowing decreases at large $p_T$ due to the increase
of $m_T$ in $y_{max}$.

In other words, due to coherence conditions, shadowing effects for
partons take place at very small $x$,
$x \ll x_{cr} = 1/m_N R_A$ where $m_N$ is the nucleon mass and $R_A$
is the radius of the nucleus.
Partons which produce
a state with transverse mass $m_T$ and a given value of Feynman $x_F$,
have $x = x_{\pm} = {1 \over 2} (\sqrt{x^2_F + 4m_T^2/s} \pm x_F)$.
Our shadowing applies to soft and hard processes. Nevertheless, for large
$p_T$ these effects are important only at very high energies, when $x \sim
{m_T \over \sqrt{s}}$ satisfies the above condition.
At fixed
initial energy $(s)$ the condition for existence of shadowing will not
be satisfied at large transverse momenta: In the central
rapidity region ($y^* =0$) at RHIC and for $p_T$ of jets (particles)
above 5(2) GeV/c the condition for shadowing is not satisfied and
these effects are absent.

So for the large $p_T$ partons it is neccessary to include the final state
interactions. The interaction of a large $p_T$ particle with
the 
medium is
described by the gain and loss differential equations which govern
final state interactions:
\beq \label{1e} {d\rho_H(x,p_T) \over d^4x} = - \widetilde{\sigma} \
\rho_S \left [ \rho_H (x, p_T) - \rho_H(x, p_T + \delta p_T)\right ]
\eeq
where $\rho_S$ and $\rho_H$ correspond to the space-time
density of the medium
and
of large $p_T$ $\pi^0$'s, $\widetilde{\sigma}$ is the interaction
cross-section averaged over momenta
and 
$d^4x$ represent the cylindrical space-time
variables: 
longitudinal proper time $\tau$, 
space-time rapidity $y$, 
and transverse coordinate $s$.

Assuming a decrease of densities with proper time
$~1/\tau$
--isoentropic
longitudinal expansion, transverse expansion is neglected--
the above equation tranforms into:
\beq
\label{2e} \tau \ {d N_{\pi^0}(b,s,y, p_T) \over d \tau} =
- \widetilde{\sigma} N(b,s,y) \left [ N_{\pi^0}(b,s,y,p_T) - N_{\pi^0}(b,
s, y, p_T + \delta p_T) \right ] 
\eeq
where $N(b,s,y) \equiv dN/dy$ $d^2s(y,b)$ is the density of the medium 
per unit rapidity and per unit of
transverse area at fixed impact parameter, integrated over $p_T$, and 
$N_{\pi^0}(b,s,y,p_T)$ is the same quantity for $\pi^0$'s at fixed
$p_T$.

If we integrate
from initial time $\tau_0$
to freeze-out time $\tau_f$ and taking into account the inverse proportionality
between proper time and densities,
$\tau_f/\tau_0 = N(b,s,y)/N_{pp}(y)$, where $N(b,s,y)$ is the density produced
in the primary
collisions in DPM
and $N_{pp}(y)$ is the density 
per unit rapidity for minimum bias
$pp$ collisions at $\sqrt{s} = 200$
GeV= 2.24 fm$^{-2}$, we obtain
the 
suppression factor $S_{\pi^0}(b,y,p_T)$ of the yield of
$\pi^0$'s at given $p_T$ and at each impact parameter, due to its
interaction with the dense medium. We get:
\beq \label{3e} S_{\pi^0}(b,y, p_T) = {\int d^2s\ \sigma_{AB}(b) \ n(b,s)
\ \widetilde{S}_{\pi^0}(b,s,y,p_T) \over \int d^2s\ \sigma_{AB}(b)\
n(b,s)} \ , \eeq
where the survival probability is given by:
\beq \label{4e} \widetilde{S}_{\pi^0}(b,s,y,p_T)  = \exp \left \{ -
\widetilde{\sigma} \left [ 1 - {N_{\pi^0}(b,s,y,p_T + \delta p_T)\over
N_{\pi^0}(b,s,y,p_T)}\right ] N(b,s,y) \ell n \left ( {N(b,s,y) \over
N_{pp}(y)}\right ) \right \}\ . \eeq
Here $\sigma_{AB}(b) = \{ 1 - \exp [-\sigma_{pp} AB \ T_{AB}(b)]\}$,
$T_{AB}(b) = \int d^2s T_A(s) T_B(b-s)$, $T_A(b)$ are the profile
functions and
$n(b,s) = AB \ \sigma_{pp}\ T_A(s) \ T_B(b-s)
/\sigma_{AB}(b)$.

One can estimate which amount of the effect takes place in the partonic phase
and which one happens in the hadronic phase.
We can divide our suppression factor in two parts:\\
Partonic, from initial density $N(b,s,y)=
{dN/dy \over \pi R_A^2} \sim {1000 \
\over \pi R_A^2} $ to ${dN/dy \over \pi R_A^2} \sim {300 \over \pi R_A^2}$,
or
equivalently from $\tau_0=1$ fm to $\tau_{p}=3.36$ fm,
and hadronic, from partonic density ${dN/dy \over \pi R_A^2} \sim {300
\over \pi R_A^2}$ to $N_{pp}(y)= {dN/dy \over \pi R^2_{pp}}
=2.24$ fm$^{-2}$, or equivalently from
$\tau_{p}=3.36$ fm to $\tau_{f}= 5-7$ fm.
We find that $75 \%$
of the effect takes place in the partonic phase while only $25 \%$ of
the effect takes place in the hadronic phase.

\section{Numerical results}
In order
to perform numerical
calculations, we need
$\widetilde{\sigma}$ --our free parameter, $\widetilde{\sigma} \sim 1$ mb-- and
the $p_T$ distribution of the $\pi^0$'s. 
We have proceeded as follows:\\
In $pp$ collisions
 at $\sqrt{s} = 200$~GeV,
the shape of the $p_T$
distribution of $\pi^0$'s can be described as
$(1 + p_T /p_0)^{-n}$
with $n = 9.99$ and $p_0 = 1.219$~GeV/c.
The corresponding
average $p_T$ is
$<p_T> = 2p_0/(n-3)$ = 0.349 GeV/c.
In $AuAu$ collisions
we assume that the $p_T$ distribution of $\pi^0$'s
at each $b$ is given by the same shape, $(1 + p_T /p_0)^{-n}$,
with
$n = 9.99$ and changing the scale $p_0$
into $p_0(b) = <p_T>_b (n-3)/2$,
where
$<p_T>_b$ is the average value
of $p_T$ measured experimentally at the corresponding centrality. For central
$AuAu$ collisions ($n_{part} = 350$
, $b=2$), $<p_T>_{exp}$ is equal to $0.453$~GeV/c resulting in a $p_0$ of
$1.583$~GeV/c.

We take $n$ as fixed, since
 at large $p_T$ the shadowing vanishes and the
ratio
$R_{AA}$ is independent of $p_T$. Also,
the experimental value of
$n$ in $dAu$ is the same as in $pp$ within errors.

We can
thus compute the ratio $R_{AA}$ in the absence of
final state interaction (colunm I, Table 1). In this case
$R_{AA}$ increases with $p_T$, there is shadowing are low $p_T$ and we have a
Cronin effect that does not decrease at large $p_T$.

To these values we apply the correction due to the suppression factor
$S_{\pi^0}$: 

First, we neglect
the second term -the gain- in eqs. (3-4) (colunm II, Table 1). We find a
general suppression in $R_{AA}$, independent of $p_T$. $R_{AA}$ increases
slightly with $p_T$ and agrees with
experimental 
data for $p_T > 5$~GeV/c. At
lower $p_T$ the result is significantly lower than the data.

\vskip 0.3cm
\begin{minipage}[t]{8.cm}
\begin{tabular}{|c|l|l|l|l|l|l|}
\hline
$p_T$ &I &II &III &IV &V &VI\\
\hline
0.5 &0.38 &0.05 &     &     &0.34 &0.38\\
2   &0.90 &0.13 &0.08 &0.11 &0.20 &0.31\\
5   &1.48 &0.21 &0.18 &0.19 &0.23 &0.25\\
7   &1.69 &0.24 &0.24 &0.24 &0.24 &0.24\\
10  &1.84 &0.27 &0.34 &0.30 &0.25 &0.24\\
\hline
\end{tabular}
\end{minipage}
\begin{minipage}[t]{7.5cm}
\vskip -1.8cm
{\small
Table 1. Values of $R_{AuAu}^{\pi^0}(p_T)$ for the
10~\% most central collisions $AuAu$ collisions at mid-rapidities
($|y^*| < 0.35$). Column I is the result obtained with no final state
interaction. The results in the other columns include final state
interaction with several ansatzs for the $p_T$ shift induced by this
interaction.}
\end{minipage}
\vskip 0.3cm
Then we introduce
the
second term -the gain- in eqs. (3-4) in two ways:

Assuming that
the $p_T$ shift of the $\pi^0$, due to its interaction with the medium,
is constant (two cases~: $\delta p_T = 0.5$~GeV/c and $\delta
p_T = 1.5$~GeV/c) (columns III and IV), we obtain a slight increase of $R_{AA}$
at large $p_T$,
rather
insensitive to the value of the shift, consistent with
data. The problem at small $p_T$ remains.

If we assume that $\delta p_T \propto (p_T - <p_T>_b)$ (columns V
and VI), in such a
way that the factor $S_{\pi^0}$ is 1 at $p_T = <p_T>$ as it should be
--no suppression at low $p_T$--, we obtain a 
slight decrease of $R_{AA}$ at large $p_T$, consistent with data and an
increase of $R_{AA}$ at low $p_T$ --due to the shift of large $p_T$ particles--
resulting in an agreement with data.

\vskip -0.5cm
\begin{flushleft}
\begin{tabular}{cc}
\psfig{figure=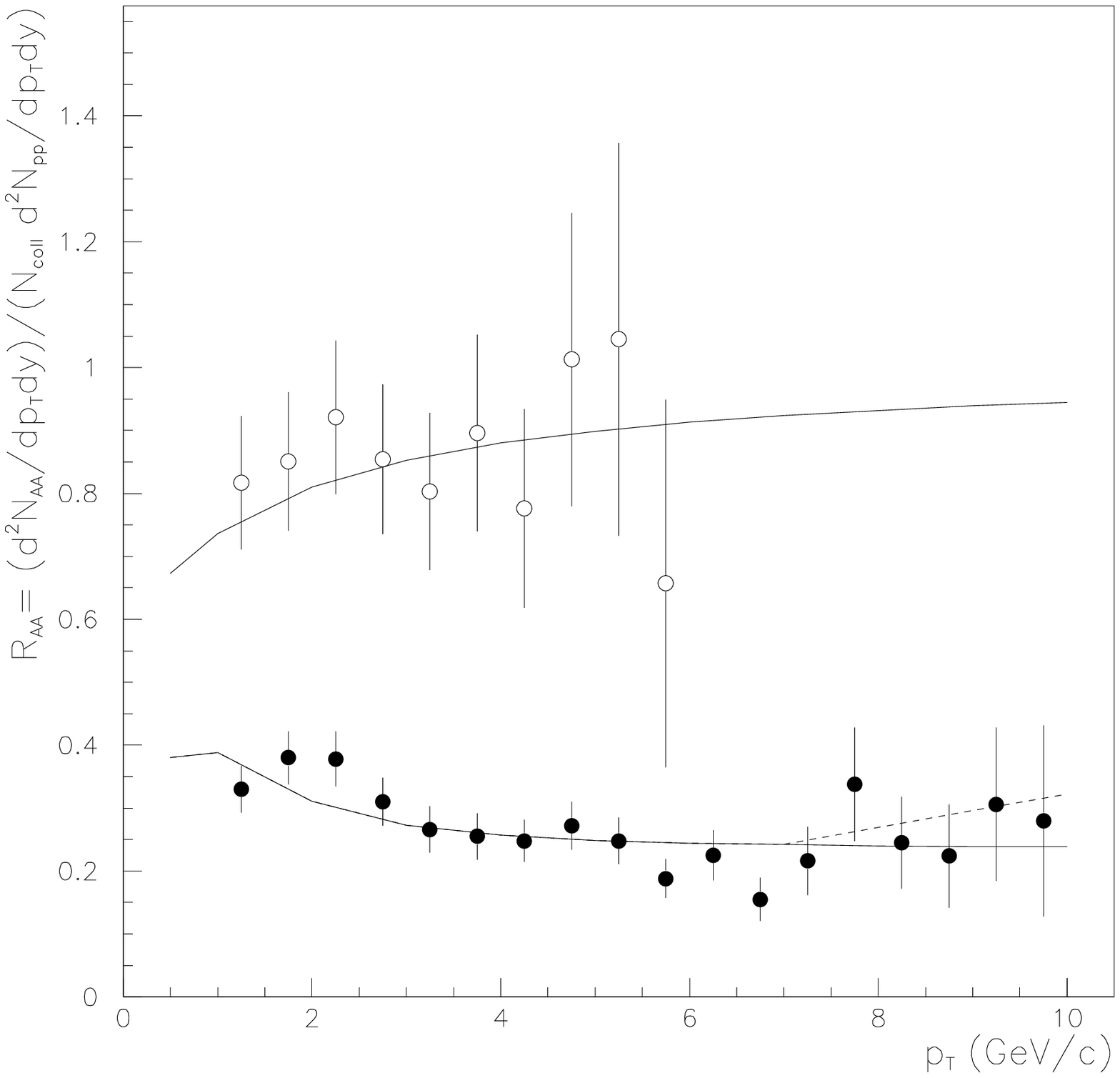,height=2.7in}
&
\psfig{figure=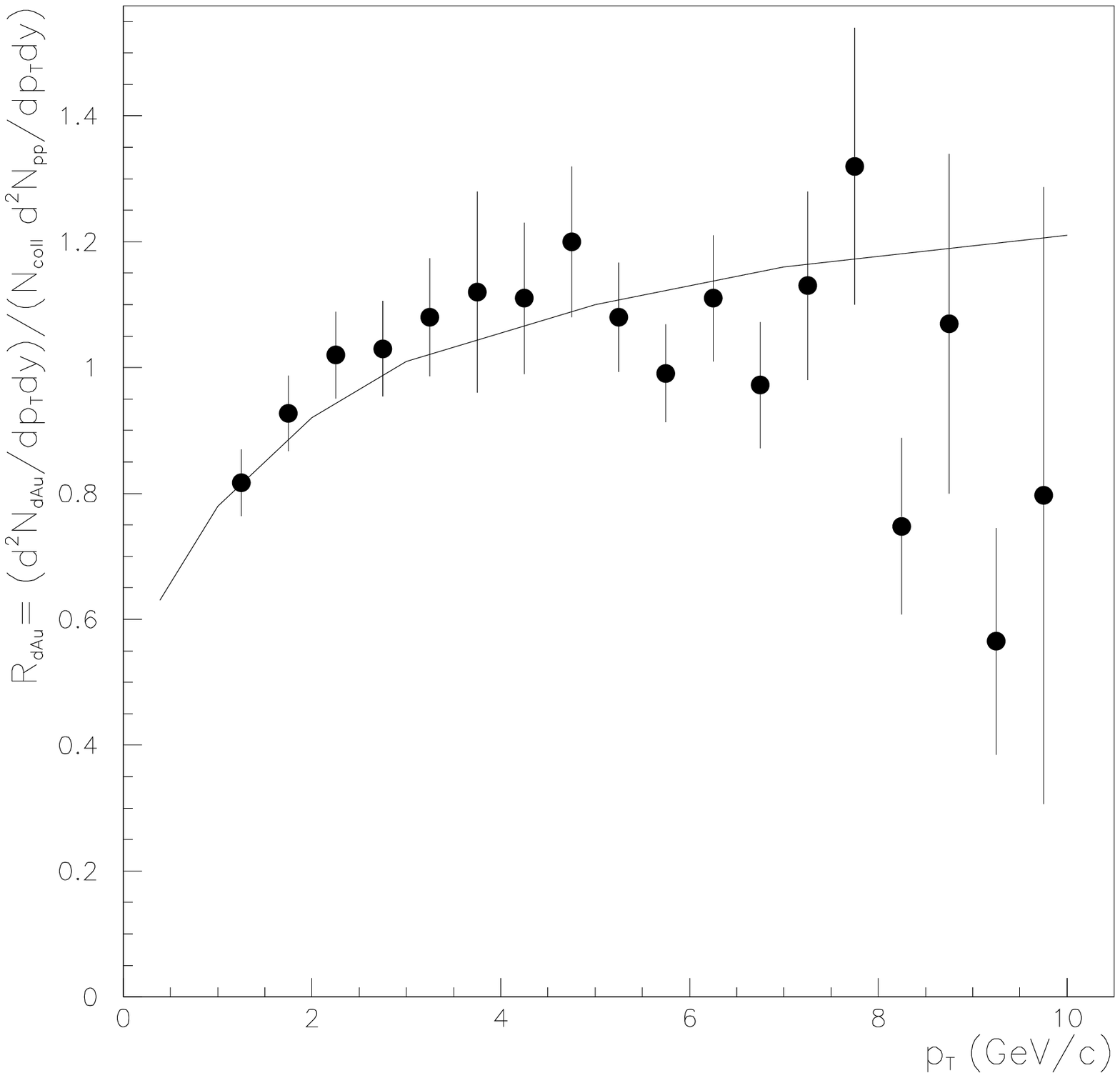,height=2.7in}\\
\end{tabular}
\end{flushleft}
\vskip -0.5cm
{\small Figure 1. Left: Values of $R_{AuAu}^{\pi^0}(p_T)$ for the
10~\% most central collisions (lower line) and for peripheral
(80-92~\%) collisions (upper line) at mid-rapidities ($|y^*| <
0.35$), using the $p_T$ shift as
$\delta p_T = (p_T - <p_T>_b)^{1.5}/20$,
 (solide line).
The dashed line is obtained using $\delta p_T = (p_T - <p_T>_b)^{1.5}/20$
 for $p_T  \leq
7$~GeV/c and $p_T$ constant for $p_T \geq 7$~GeV/c. Right: 
Values of $R_{dAu}^{\pi^0}(p_T)$ for minimum bias collsions at
mid-rapidities ($|y^*| <
0.35$), using the $p_T$ shift as
$\delta p_T = (p_T - <p_T>_b)^{1.5}/20$. The data are from PHENIX \cite{3r}.}
\vskip 0.3cm
We turn next to minimum-bias $dAu$ collisions at central rapidity. Here $<p_T> =
0.39$~GeV/c. With the same value of $n$ as above $(n
=9.99)$, this corresponds to $p_0 = 1.346$. Calculating the ratio
$dAu$ to $pp$
we obtain the result of Fig. 1. 
At forward
rapidity, $R_{dAu}$ decreases as $y$ increases due two effects:\\
The first effect is basically due to
energy-momentum conservation. It has been known for a long time in
hadron-nucleus collisions at SPS energies --low $p_T$ ``triangle''--.
Its extreme form occurs in the hadron fragmentation region,
where the yield of secondaries in collisions off a heavy nucleus is
smaller than the corresponding yield in hadron-proton. This phenomenon
is known as nuclear attenuation. It turns out, that, at RHIC energies,
this effect produces a decrease of $R_{dAu}$ of about 30~\% between
$\eta^* = 0$ and $\eta^* = 3.2$.\\ 
The second effect is the increase of the shadowing corrections in $dAu$
with increasing $y^*$.  This produces a decrease of
$R_{dAu}(p_T)$ between $\eta^* = 0$ and $\eta^* = 3.2$ of about 30~\%
for pions produced in minimum bias collisions. Therefore, we expect a
suppression factor of about 1.7 between $R_{dAu}(p_T)$ at $\eta^* = 0$
and at $\eta^* = 3.2$, practically independent of $p_T$. This is
consistent with the BRAHMS results. 

\end{document}